\documentclass[
               journal=jpclcd,
               manuscript=article
               ]
               {achemso}

\setkeys{acs}{articletitle = true}

\usepackage[version=3]{mhchem} 
\usepackage{xcolor}
\usepackage{graphicx}
\usepackage{mciteplus}

\newcommand*\oriol[1]{\textcolor{red}{#1}}

\newcommand*\change[1]{\textcolor{blue}{#1}}

\author{Jing Sun}
\email{jing.sun@pci.uni-heidelberg.de}
\affiliation[Heidelberg University]
{Theoretische Chemie, Physikalisch-Chemisches Institut,
Universität Heidelberg,\\ 69120 Heidelberg, Germany}
\author{Oriol Vendrell}
\email{oriol.vendrell@uni-heidelberg.de}
\affiliation[Heidelberg University]
{Theoretische Chemie, Physikalisch-Chemisches Institut,
Universität Heidelberg,\\ 69120 Heidelberg, Germany}

\title[Chemical rates in cavities]
    {Modification of Thermal Chemical Rates in a Cavity via
    Resonant Effects in the Collective Regime}

\begin{document}

\begin{tocentry}
    \includegraphics[width=5cm]{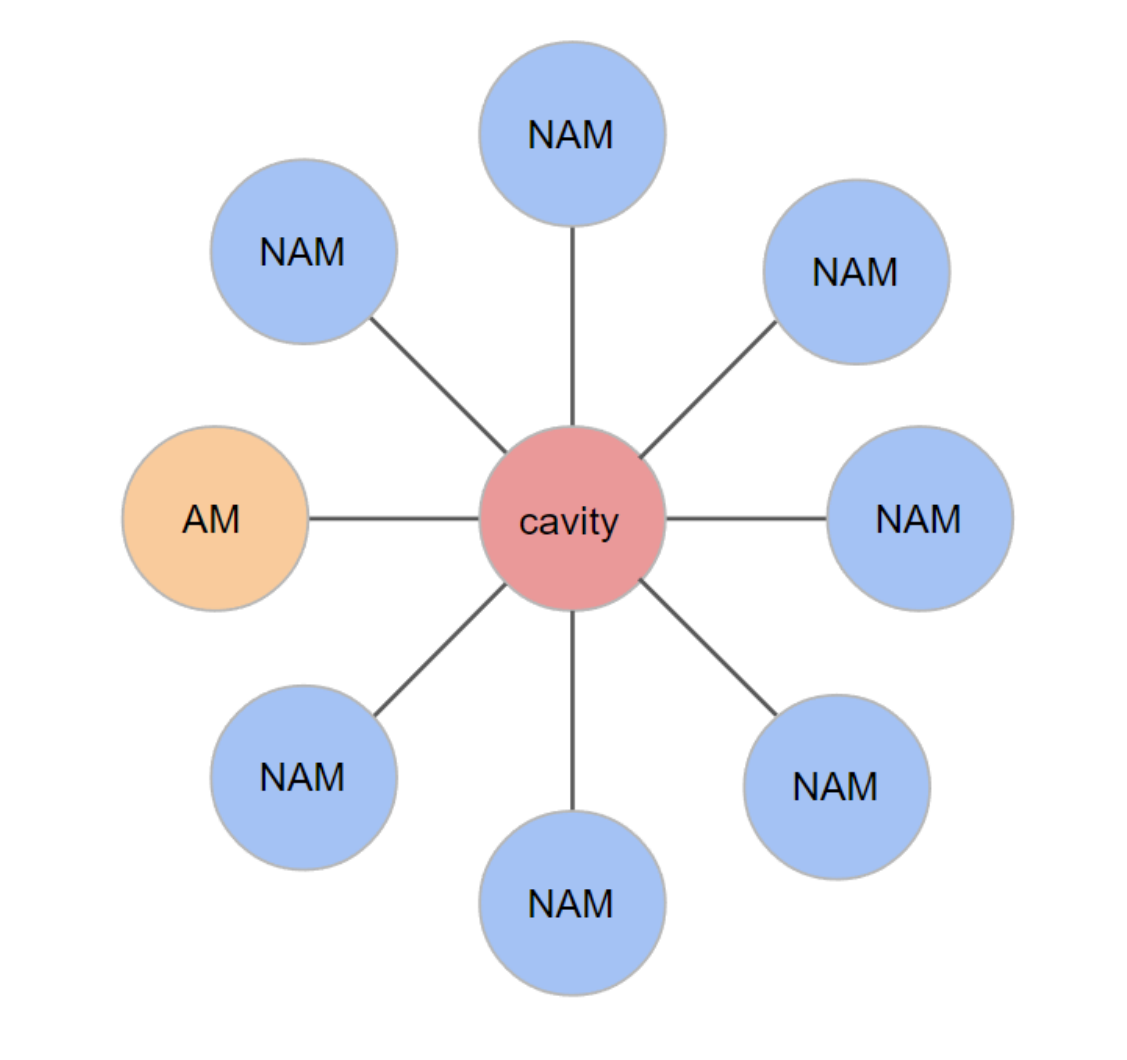}
\end{tocentry}




\begin{abstract}

  The modification of thermal chemical rates in Fabry-Perot cavities,
  as observed in experiments, still poses theoretical challenges.
  While we have a better grasp of how the reactivity of isolated molecules
  and model systems changes under strong coupling, we lack a comprehensive
  understanding of the combined effects and the specific roles played
  by activated and spectator molecules during reactive events.
  In this study, we investigate an ensemble of randomly oriented gas-phase
  HONO molecules undergoing a \emph{cis-trans} isomerization reaction on
  an \emph{ab-initio} potential energy surface.
  Using the classical
  reactive flux method, we analyze the transmission coefficient and
  determine conditions that lead to accelerated rates within the
  collective regime.
  We identify two main mechanisms at work: firstly, spectator molecules
  enhance the cavity's ability to dissipate excess energy from the activated
  molecule post-reactive event.
  Additionally, the interaction between spectator molecules and the cavity
  gives rise to the creation of polaritonic modes.
  These modes then interact with the activated molecule at a
  shifted resonance frequency.
\end{abstract}



Vibrational strong coupling (VSC) is an active area of research within the field of
polaritonic chemistry but the effect of Fabry-Perot cavities on molecular reactivity
is still a poorly understood phenomenon.
Under VSC, Fabry-Perot cavities can resonantly couple to the infrared active modes of molecules
thus leading to Rabi splitting and the formation of vibrational polaritonic bands under infrared
irradiation~\cite{hut_12_1624,sha_15_5981,ebb_16_2403,dun_16_13504,dun_18_965,yan_20_919,fas_21_11444}.
Additionally, an increasing number of experimental works have reported the modification
of the rate of thermal chemical reactions inside the cavity in the dark, i.e. under
coupling to background thermal radiation and vacuum fluctuations of the confined
electromagnetic
fields~\cite{tho_16_11462,lat_19_10635,tho_19_615,ver_19_15324,tho_20_249,imp_21_191103,Won_23_6650}.
These experimental results have prompted the development of various
theoretical models to explain how cavities alter the ground electronic state
structure and spectroscopy, and more recently, how they modify reaction
rates~\cite{li_21_094124,li_21_1315,li_21_15533,yan_21_9531,man_22_014101,lat_19_10635,du_22_096001,tho_19_615,wan_22_3317}.
These models often consider a single molecule coupled to a cavity mode whereas,
in strong contrast,
$10^6$-$10^{12}$ molecules are collectively coupled to the cavity
in actual Fabry-Perot resonators~\cite{pin_15_053040}.
As a result, in order to recreate the same Rabi splitting and overall light-matter
coupling as in experiments, theoretical models must consider a single or a few
molecules in the strong to ultra-strong coupling regime.
In terms of chemical kinetics, closing gap between multiple molecules,
each weakly interacting with the cavity,
and a single strongly-coupled molecule remains an
open question, although theoretical proposals that cleverly exploit the
high symmetry in the Hamiltonian have been put forward recently~\cite{per_23_e2219223120}.

Early attempts to understand the effect of cavities on thermal reactivity were based
on transition state theories under the assumption that the main effects correspond
to modifications of the potential energy barrier separating reactants and products.~\cite{li_20_234107,cam_20_161101}
While this is conceivable in the single-molecule,
strong to ultra-strong coupling regime,
alterations to the potential energy barrier of individual molecules
cannot play a significant role in the
\emph{collective} strong coupling regime.
More recently, the idea that the cavity effects must be of a dynamic and more
subtle nature has gained traction~\cite{jin_22_4441,Lin_22_6580,wan_22_3317}.
Accordingly, the cavity modifies the energy
redistribution pathways of the reacting molecules, thus facilitating or hindering
the reactive process.
Dynamical effects are comparatively small and can be captured as
a correction factor to transition state theories.
This is in line with the experimental findings,
which consistently demonstrate that the most substantial alterations in chemical rates reported
thus far fall within a range of no more than one to two orders of magnitude
(see e.g. Table 1 in Ref. \citenum{nag_21_16877} and
subsequent references).

In previous work, we studied the isomerization reaction of the HONO molecule
inside a cavity at room temperature under fixed orientation~\cite{jin_22_4441}.
We established
that the cavity affects the energy redistribution to and from the reactive
coordinate during the reactive event, and that this effect occurs when the
cavity couples to infrared-active molecular coordinates directly or indirectly
involved in the reactive process. This results in relatively sharp
modifications of the rate as a function of the cavity frequency.
Full quantum calculations on model systems have also established sharp
modifications of the reaction rate when the cavity is resonantly coupled to the
reaction coordinate~\cite{Lin_22_6580}.
To the best of
our knowledge, the simulations in
Ref.~\citenum{jin_22_4441} contained the
first converged canonical
reaction rate calculations
an anharmonic \emph{ab intio} potential energy surface
with full
account of dynamical effects that were conducted for a molecule coupled to
a cavity.
From a detailed mechanistic perspective, there still remains the open question
of how the molecular ensemble, together with the cavity, participates in the
modification of the reaction rate.
Notably, recent work based on a three-dimensional reactive model with up to
several thousand molecules finds that the microcanonical survival probability of
a unimolecular dissociation can be modified collectively by the cavity when it
becomes resonant with the vibrational modes of the molecules, and it indicated
that the spectator molecules participate collectively.~\cite{wan_22_3317}

In this work,
we consider ensembles of freely
rotating HONO molecules in the gas phase and explore the collective effects on
the \emph{cis}-\emph{trans} isomerization reaction rate.
The single-molecule coupling strength is chosen small enough such that an effect
on the reaction rate is still seen for one single molecule, but such that the
Rabi splitting is still small. The latter means that, under irradiation, no
defined polaritonic bands are present.
Under such conditions, we explore the situation in which the addition of
molecules at a constant coupling strength results in the appearance of
vibropolaritonic bands.
We can demonstrate numerically that, in this regime, the resonance of the single
activated molecule with the polaritonic bands of the cavity-ensemble system
results in the same effect as the direct coupling of a single molecule
with the cavity mode.
%
Moreover, we show in detail how theorientation of the activated and
non-activated molecules may differ for optimal interaction with the cavity, and
the resonance may frequency may shift for the activated molecule due to its
higher energy content.

%
%
Although the size of the considered molecular ensembles is still orders of
magnitude away from the actual experiments in Fabri-Perot cavities, we choose an
interaction strength that brings the system from a weak to a collectively strong
coupling as a function of the ensemble size. 
Our simulations demonstrate the collective effect on chemical rates in the
under-damped regime and explain how the cavity enhances the rate when more
molecules are added to the system. 
Finally, key to our analysis is the fact that the single activated molecule
(AM), the molecule undergoing the chemical reaction at a specific moment in
time, and the $(N-1)$ spectator or non-activated molecules (NAM) play
fundamentally different roles in the collective mechanism.
%

%

%
Here, we briefly introduce the Hamiltonian of the molecular ensemble on non-interacting
gas-phase molecules coupled to a single cavity mode
%
%
%
\begin{align}
    \label{eq:Ham}
    {H} & = \sum_{l=1}^{N}{H}_{mol}^{(l)} + {H}_{cav}
\end{align}
where the molecular and cavity terms read
\begin{align}
    \label{eq:HamMol}
    {H}_{mol}^{(l)} & = \sum_{j_l=1}^{N_a} \frac{ {{\bf P}_{j}^2}^{(l)} } {2M_{j}}
     + {V}({\bf X}_{1}^{(l)}\ldots {\bf X}_{N_a}^{(l)}), \\
    \label{eq:HamCav}
    {H}_{cav} &= \frac{1}{2}\left[ {p}_{cav}^2 +\omega_{cav}^2
    \left( {q}_{cav}+\frac{\boldsymbol{\lambda}}{\omega_{cav}}
    \cdot
    \sum^{N}_{l=1}{\boldsymbol{\mu}}^{(l)})\right)^2 \right].
\end{align}
Here $V({\bf X}_{1}^{(l)}\ldots {\bf X}_{N_a}^{(l)})$ denotes the ground electronic state potential energy
surface (PES) of the $l$-th molecule with Cartesian
positions ${\bf X}_{j}^{(l)}$ and momenta ${\bf P}_{j}^{(l)}$.
The \emph{ab initio} PES was obtained by Richter et al.~\cite{ric_04_1306}
at the CCSD(T) level of theory.
Hence, each $l$-th molecule is considered in its ground electronic state
under the Born-Oppenheimer (BO) approximation and interacting with the
cavity via its field-free dipole-moment vector surfaces
$\boldsymbol{\mu}^{(l)}\equiv \boldsymbol{\mu}^{(l)}({\bf X}_{1}^{(l)}\ldots {\bf X}_{N_a}^{(l)})$.
Similarly to other studies and to facilitate comparisons, we introduce
the coupling parameter $g = \lambda \sqrt{\hbar\omega_{cav}/2}$, which has units
of electric field.

\begin{figure}[t]
\includegraphics[width=7cm]{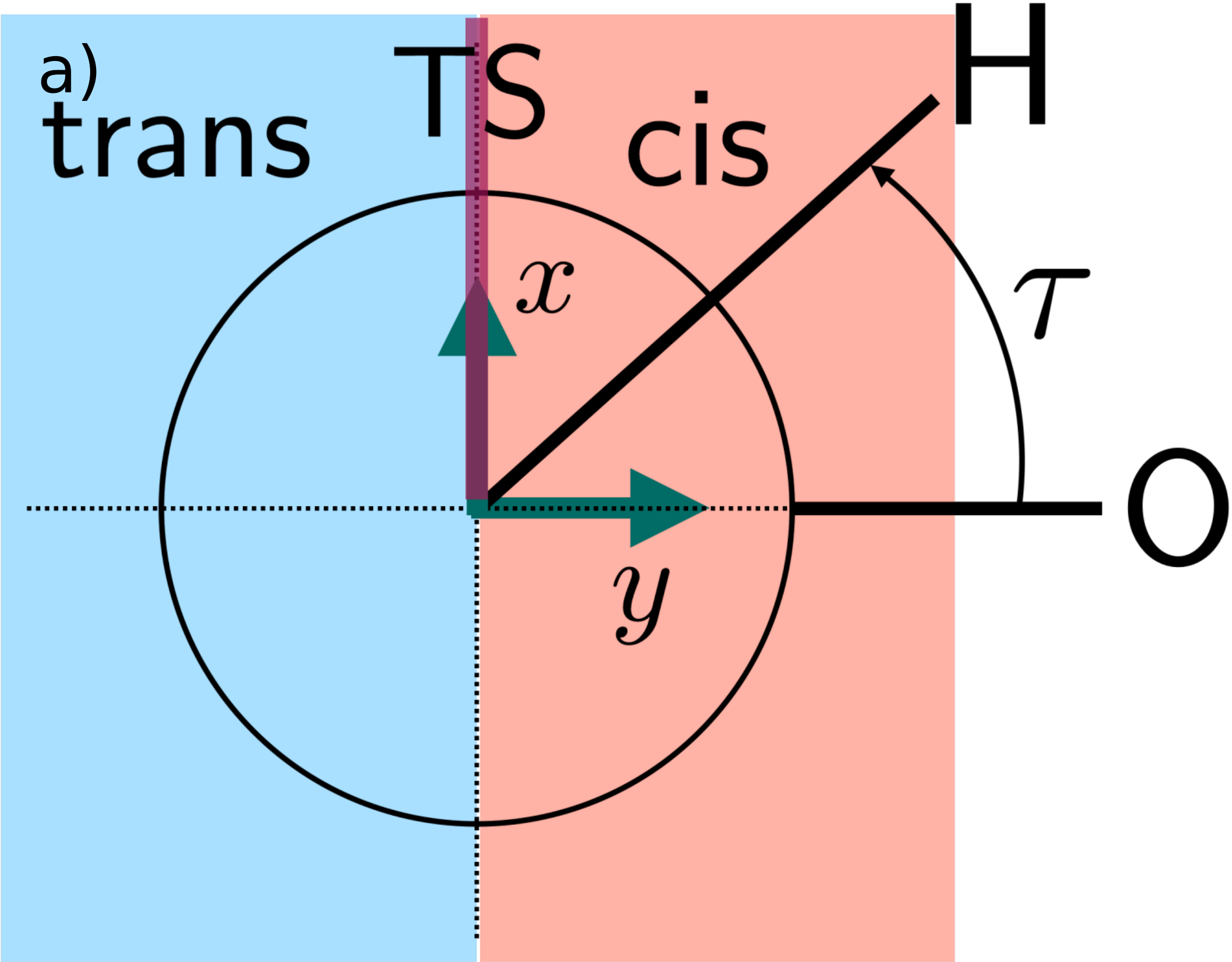}
\includegraphics[width=8cm]{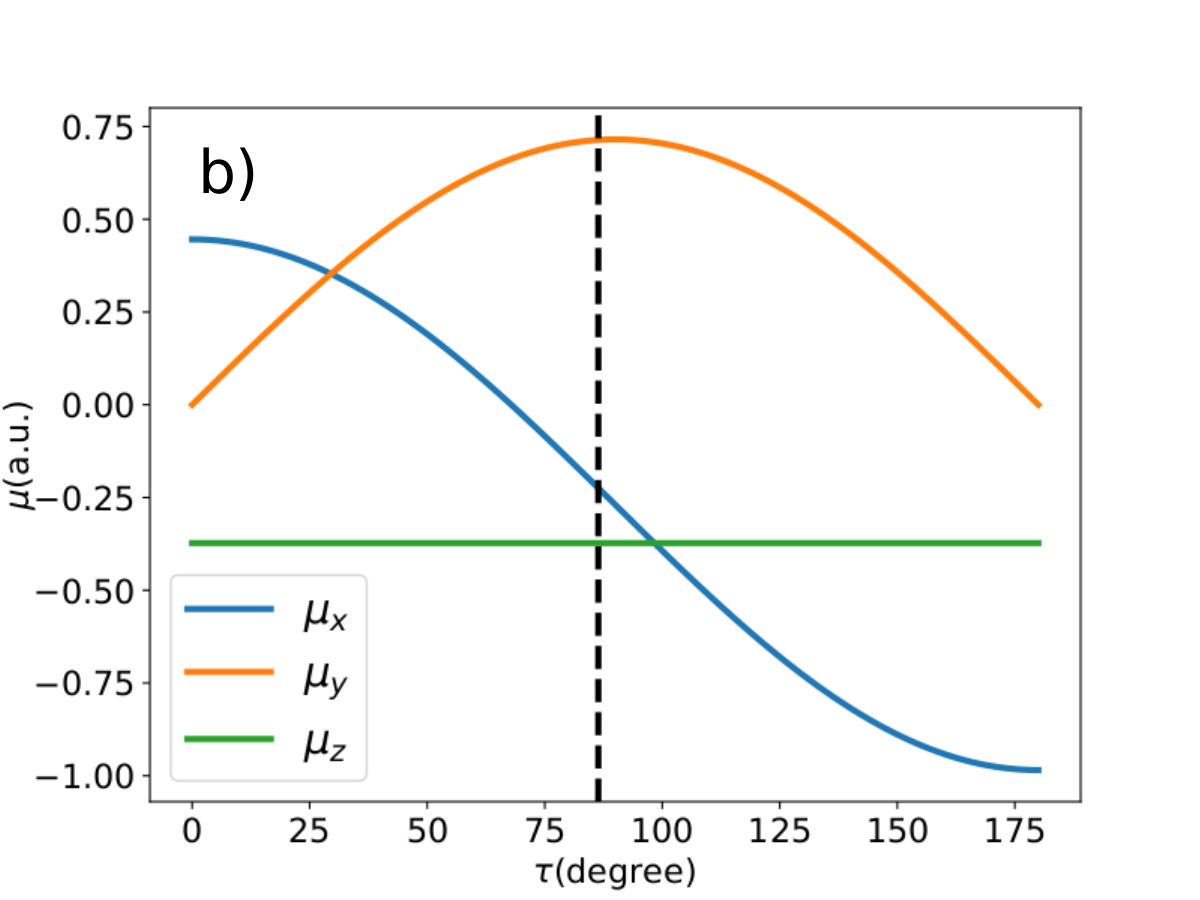}
\caption{
    \label{fig:newman}
    {
    a) Newman diagram of the HONO molecule showing the definition of the
    body-fixed axes. The hydrogen atom is on the front
    side, the terminal oxygen atom is on the back, and the remaining oxygen and
    nitrogen atoms lie at the origin of the diagram along the perpendicular
    $z$-axis.
    The $(y,z)$-plane is determined by the ONO atoms, and the reaction
    coordinate $\tau$ corresponds to the dihedral rotation of the H-atom
    around the $z$-axis and on the $(x,y)$-plane. Its origin is set at the
    minimum energy configuration of the \emph{cis} region.
    The transition state lies at about 90 and by symmetry also at about 270 degrees.
    b) Permanent dipole of the HONO molecule in atomic units as a function of
    the torsion coordinate $\tau$.  The axes are referred to the molecular
    frame axis~\cite{jin_22_4441}. The black dashed line represents TS.
    }
}
\end{figure}

According to the reactive flux method for the classical rate
constant~\cite{mon_79_4056,ros_80_162,cha_87_,ber_88_3711,kuh_88_3261}, the
\emph{cis}-\emph{trans} reaction rate is described as
\begin{align}
    \label{eq:kt}
    K(t) &= x_{cis}^{-1}
    \langle \dot{\tau}(0)\,
            \delta[\tau(0)-\tau^{\ddag}]\,
            \theta[\tau(t)] \rangle,
\end{align}
where $x_{cis}$ is the equilibrium fraction of HONO at the \emph{cis} geometry,
$\dot{\tau}(0)$ is the initial velocity of a phase-space point perpendicular to
the dividing surface between reactants and products, and $\tau^{\ddag}$ is the
torsion angle corresponding to the transition state (TS) geometry.
A corresponding diagram illustrating the reaction coordinate and the \emph{cis} and
\emph{trans} configurations is found in Fig.~\ref{fig:newman} in the form
of a Newman diagram with the central N-O bond perpendicular to the plane of the
paper and at the origin of the $(x,y)$-plane.
The Heaviside function $\theta[\tau]$ is one for the \emph{trans}
configurations, and zero otherwise.
The brackets represent the thermal ensemble average
over all trajectories, where a temperature of 300~K is considered for all
calculations.
In practice, one obtains the reactive flux when $K(t)$
reaches its plateau value~\cite{ber_88_3711}. In cases where a plateau value
cannot be reached, for example for two potential energy minima
separated by a one-dimensional barrier without any dissipation, a reaction
rate cannot be defined in the sense of Eq.~\ref{eq:kt}.
Transmission coefficients and rate constants without a time argument
refer to their plateau value, and
$\lim_{t\to 0^{+}}K(t) = K_{TST}$,
where $K_{TST}$ can also be evaluated through Eyring's
equation~\cite{eyr_35_107,han_90_251}.
$K(t)$ can be related to $K_{TST}$ through
the introduction of a transmission coefficient $\kappa(t)$,
$K(t) = \kappa(t) K_{TST}$,
where $\kappa(t)$ is in practice the quantity of interest obtained from classical
trajectories~\cite{ber_88_3711}.
The enhancement or suppression of the reaction rate by the cavity
can then be quantified through the comparison of the rate constant
inside ($K^{(c)}$) and outside the cavity ($K^{(0)}$).
Now, as long as the dividing surface between reactants and products is
chosen to lie perpendicular to the cavity coordinate (cf. Fig. 3 in Ref.~\citenum{jin_22_4441}), i.e.
the same dividing surface is used to define the TS with and without cavity,
one can rely on the very good approximation that $K_{TST}^{(c)} \approx K_{TST}^{(0)}$~\cite{jin_22_4441}.
As a consequence, the cavity effect on the rate follows directly from the ratio
of the transmission coefficients
\begin{align}
    \label{eq:prodkappa}
    R = \frac{\kappa^{(c)}}{\kappa^{(0)}}.
\end{align}
Although $\kappa^{(c)}$ and $\kappa^{(0)}$ lie in the $[0,1]$ range,
the ratio $R$ indicating the chemical rate enhancement or suppression can, in principle,
be either larger or smaller than one.

%
Finally, it is important to define how to calculate the reactive flux with the
expression in Eq.~\ref{eq:kt} when considering a molecular ensemble, since all
coordinates of all molecules enter in the definition of the various quantities.
In doing so, it is essential to realize that the fraction of activated molecules
(AM) $F_{AM}$, those which are undergoing the chemical transformation at a
specific moment in time, to non-activated molecules (NAM) is in general very
small.
For a unimolecular reaction with a forward rate constant $K$ and a total of $N$
molecules coupled to the cavity, the rate of molecules that start the
transformation per unit of time is $dN/dt = KN$.
Multiplying by $\Delta_{TS}$, the amount of time the molecule spends crossing
the transition state (TS), and dividing by the total number of molecules $N$
results in an estimate for $F_{AM} = K \Delta_{TS}$.
Considering a rate of about $10^{-4}$~s$^{-1}$ as in the example of
Ref.~\citenum{ebb_16_2403}, and with $\Delta_{TS}\approx 10^{-13}$~s for a
typical reaction in solution~\cite{ger_91_74}, one obtains $F_{AM} \approx
10^{-17}$. A rate of $10^{4}$~s$^{-1}$, of the order of the HONO isomerization
studied here, results in $F_{AM} \approx 10^{-9}$, still negligible compared to
the fraction of spectator NAMs.
Hence, when considering the modification of chemical rates for ensembles under
strong coupling, one must assume that on average only one AM crosses the barrier
at a time.
In our classical treatment of the reactive flux, this means that the sampling of
transition state configurations according to Eq~\ref{eq:kt} is performed for one
molecule only, while the other molecules are in thermal equilibrium close to
their minimum energy configurations in the reactant potential energy well.
Every set of different model parameters of the simulations studied below consists
of a batch of $10^{4}$ trajectories 
sampled from a canonical ensemble. The simulations are built on top of the
OpenMM package for customizable molecular simulation.~\cite{Eas_17_59}
%



The three spatial directions of the dipole in the body-fixed molecular frame
along the reaction coordinate $\tau$ are shown in Fig~\ref{fig:newman}b.  The
modulation of the $\mu_z$ component is very small as a function of $\tau$ and
its largest modulation occurs instead in the stretching and bending modes of the
molecular skeleton. The $\mu_z$ component of the dipole is the only one
available from \emph{ab initio} calculations. Hence, to be consistent, here we
use a simple partial charge model of the dipole for its three components that
agrees qualitatively well with calculated cuts of the \emph{ab initio} dipole
surfaces. Details are found in the supporting material.
%

\begin{figure}[t]
\includegraphics[width=8cm]{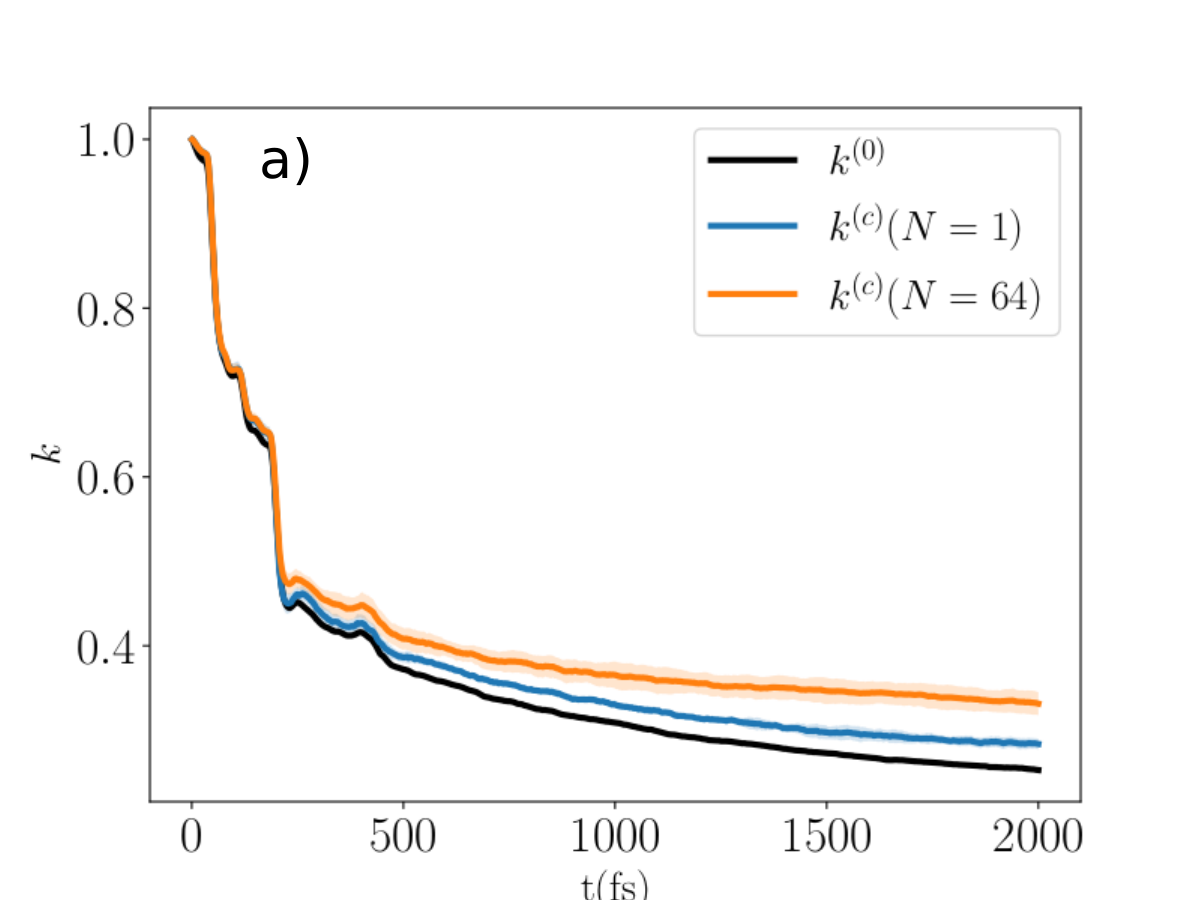}
\includegraphics[width=8cm]{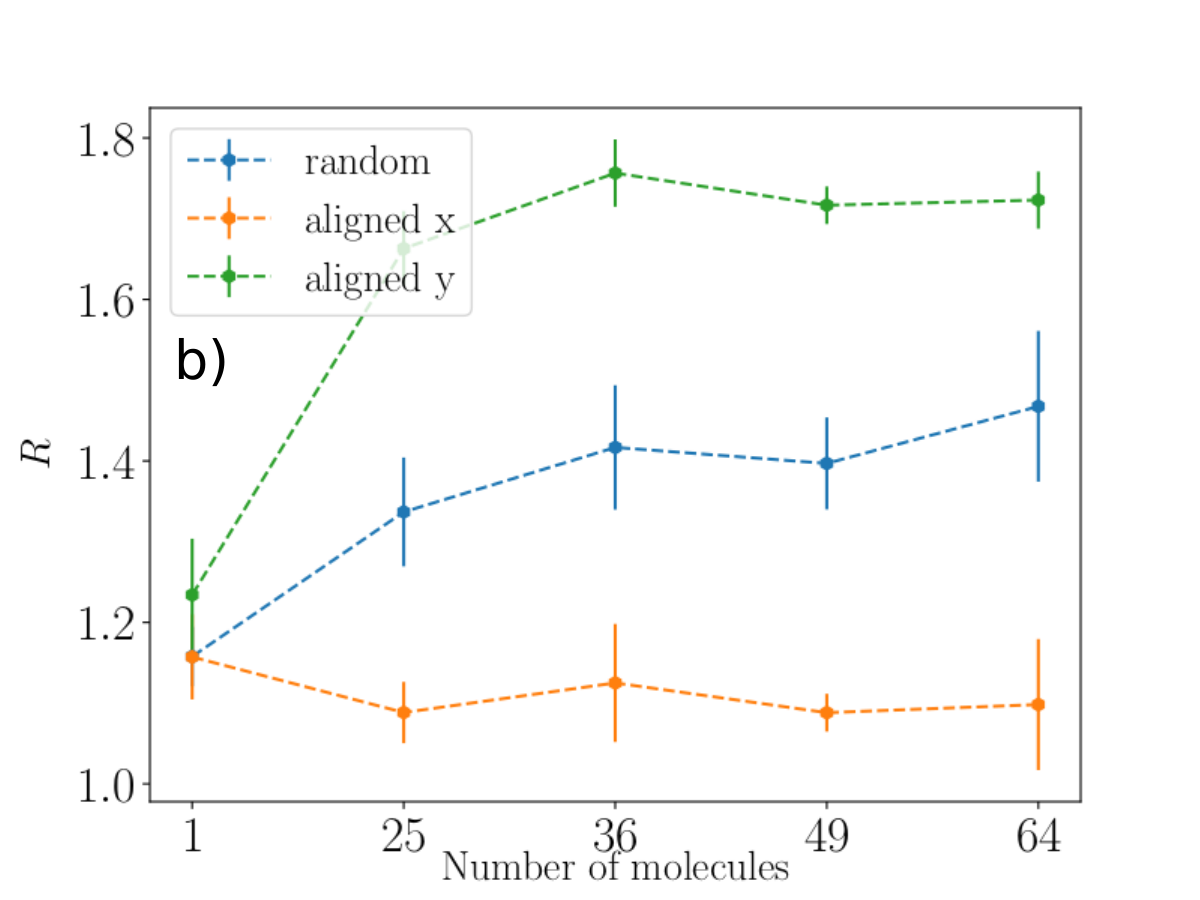}
\caption{
    \label{fig:R_collective}
    a) Transmission coefficient $\kappa^{(c)}(t)$ with fixed $g=0.4$ (V/nm) with
    different $N_{mol}$, $\omega_{cav}=640$~cm$-1$ b) $R$ for increasing number
    of molecules with fixed $g=0.4$ V/nm (see text for details). Random case
    means that all molecules always randomly orientate when evolution. Aligned
    $x$($y$) molecules mean that the cavity polarization is aligned with HONO's
    $x$($y$)-axis and orientations of all HONO are fixed at all time. 
    }
\end{figure}

We begin the discussion of the results by considering the effect of the cavity
on an ensemble of $N$ randomly oriented HONO molecules for
$\omega_{cav} = 640$~cm$^{-1}$, which is the frequency of the reaction coordinate
at the minimum-energy geometry of the \emph{cis} potential well~\cite{jin_22_4441}.
Fig~\ref{fig:R_collective}a presents the transmission coefficient as a function
of $N$ for randomly oriented HONO molecules.
Outside the cavity, $\kappa^{(0)} \approx 0.23$ at 300 K as can be seen in the
black trace, Fig~\ref{fig:R_collective}a, and also reported in
Ref.~\citenum{jin_22_4441}.
This relatively low transmission is caused by a slow rate of intra-molecular
vibrational energy redistribution (IVR) from the reaction coordinate of the AM
to the rest of the molecular system in the low friction regime.
Now, for a single molecule ($N=1$) inside the cavity we fix the coupling
strength to $g=0.4$ V/nm, smaller than the weakest coupling considered in
Ref.~\citenum{jin_22_4441}, but still showing a clear enhancement of the
chemical rate in line with the results reported there.  Note,
however, that now the single molecule is randomly oriented and its atomistic
trajectories in Cartesian space include rotations and vibrations.
Next, we consider $N=64$, where one molecule corresponds to the AM while $N-1$
are spectator NAMs.
Rather than fixing the macroscopic Rabi splitting by multiplying
the coupling strength with the $1/\sqrt{N}$ factor, we keep it constant to
$g=0.4$~V/nm.
Hence, the AM is coupled to the cavity with the same coupling strength in
both the $N=1$ and the $N=64$ cases.
This way, the modification of the rate cannot be attributed to an artificial
re-scaling of the cavity to single-molecule coupling, and can only be attributed
to a genuine collective modification of the dynamics of the AM through the
presence of spectators molecules. The latter are only indirectly coupled to
the AM via the cavity mode. 
For $N=64$, $\kappa^{(c)}$ is further enhanced, as seen by comparing the orange
and blue curves in Fig~\ref{fig:R_collective}a.  The cavity plus NAMs further
accelerate the chemical reaction by increasing the total transmission
coefficient compared to $\kappa^{(0)}$ and to $\kappa^{(c)}(N=1)$ i.e. $R > 1$.

In the following, we describe the mechanistic insights behind this numerical
experiment. 
First of all, we must note that the reported effect
stagnates at $N\approx 36$ for the choice of model parameters when considering
$R(N)$ in Fig.~\ref{fig:R_collective}b for the randomly oriented molecules.
As such, the reported rate modification does not explain the macroscopically~\cite{du_22_096001}
large $N$ limit, which remains a standing unresolved
issue in the field, but which is not the focus of this contribution.
We emphasize that it is not the goal of this paper to address the macroscopic
large $N$ limit involving correspondingly small single-molecule couplings.
Indeed, in the opinion of the authors, this limit likely escapes the framework
provided by Hamiltonian~\ref{eq:Ham}, irrespective of whether it is considered
quantum mechanically or by classical mechanics.

However, our simulations shed important light on the distinct roles played by
the AM and NAMs in the few molecules, vibrational strong coupling regime.
Thus, in the following our goal is to explain the mechanisms by which the added
NAMs modify the reaction rate of the AM, even though the AM and NAMs are not
directly coupled and only interact with each other indirectly through the cavity degree of
freedom.

Let us first compare the simulations with randomly oriented initial conditions
to simulations performed with aligned molecules.  
When all molecules have their $x$-axis (cf. Fig.~\ref{fig:newman}) aligned
with the cavity polarization the collective effect disappears, as
illustrated by the orange trace in Fig~\ref{fig:R_collective}b.
In stark contrast, when the cavity
polarization is aligned with HONO's $y$-axis, the enhancement of the reaction
rate with increasing $N$ is even more pronounced as compared with the randomly oriented
case.
Comparing the three cases, $R$ is similar when $N=1$ and
a modification of the rate in the $N>1$ case only occurs
for randomly oriented and $y$-aligned cases.
This observation indicates that the orientation of the molecules with
respect to the cavity polarization can be very important, and that the
AM and the $N-1$ NAM (spectators) play different roles in connection with their
orientation inside the cavity.
Indeed, molecules with their $x$-axis oriented with the cavity polarization experience
the largest modification of the permanent dipole as they cross the transition
state region, blue trace in Fig.~\ref{fig:newman}b. Therefore, the AM efficiently couples
with the cavity in this orientation. The AM also couples efficiently with the cavity
when its $y$-axis is oriented with the cavity polarization. Here, the most efficient coupling
will occur when the AM visits the \emph{cis} or \emph{trans} regions of the configurational
space (see orange trace in Fig.~\ref{fig:newman}b), which occurs just a few femtoseconds
before or after having passed the TS.
This situation is radically different for the NAMs because their energy content along the
reaction coordinate is thermal and much lower than the AM. Thus, they are confined to oscillate
close to the potential energy minimum, where, in the case of HONO,
$\partial\mu_x/\partial\tau=0$.
Thus, when the cavity is tuned to the frequency of the HONO torsion, only $y$-aligned NAMs
can participate, whereas $x$-aligned NAMs are literally invisible to the cavity.
Summarizing, both the AM and the $N-1$ NAM must be efficiently coupled to the cavity for
collective effects to take place,
which may not necessarily occur for the same orientation,
or even cavity frequency, as we illustrate below.

%
\begin{figure}[t] \includegraphics[width=8cm]{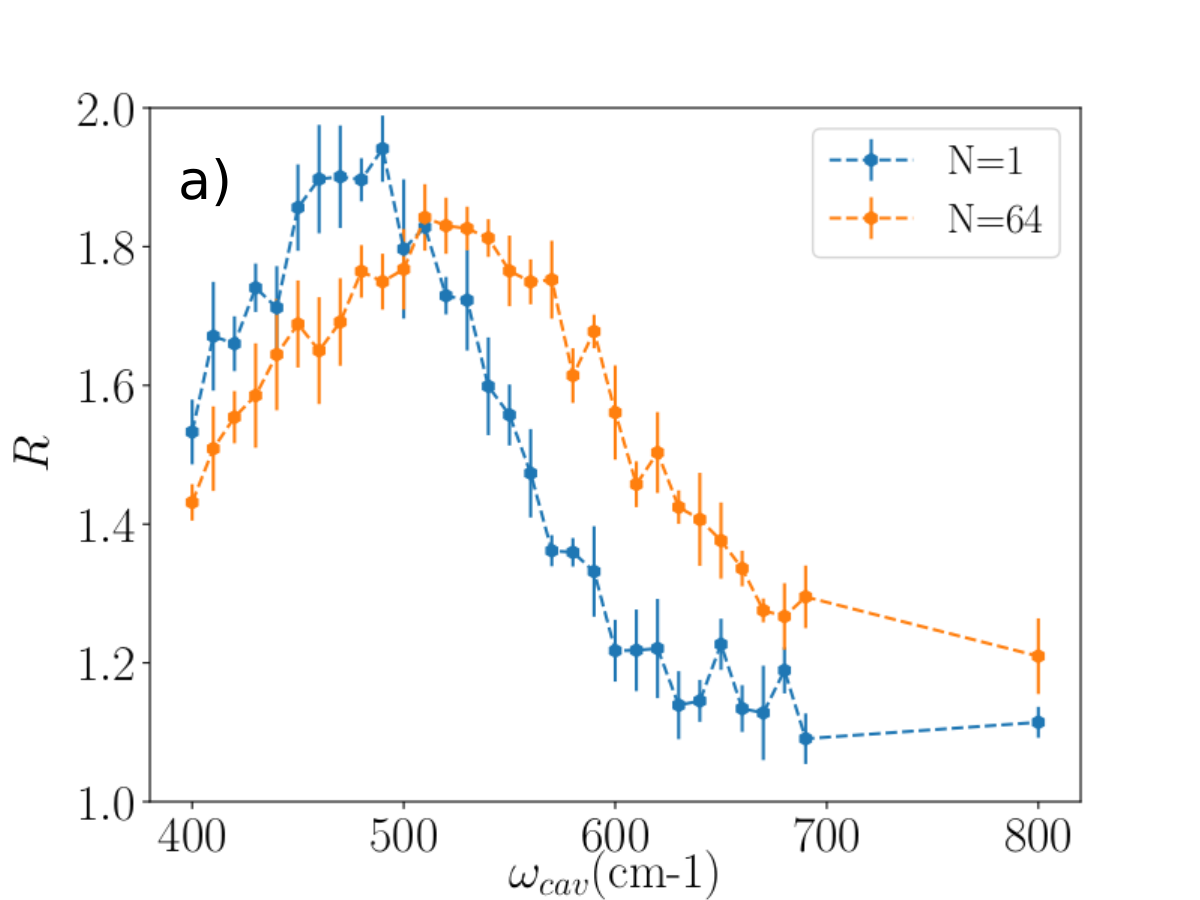}
    \includegraphics[width=8cm]{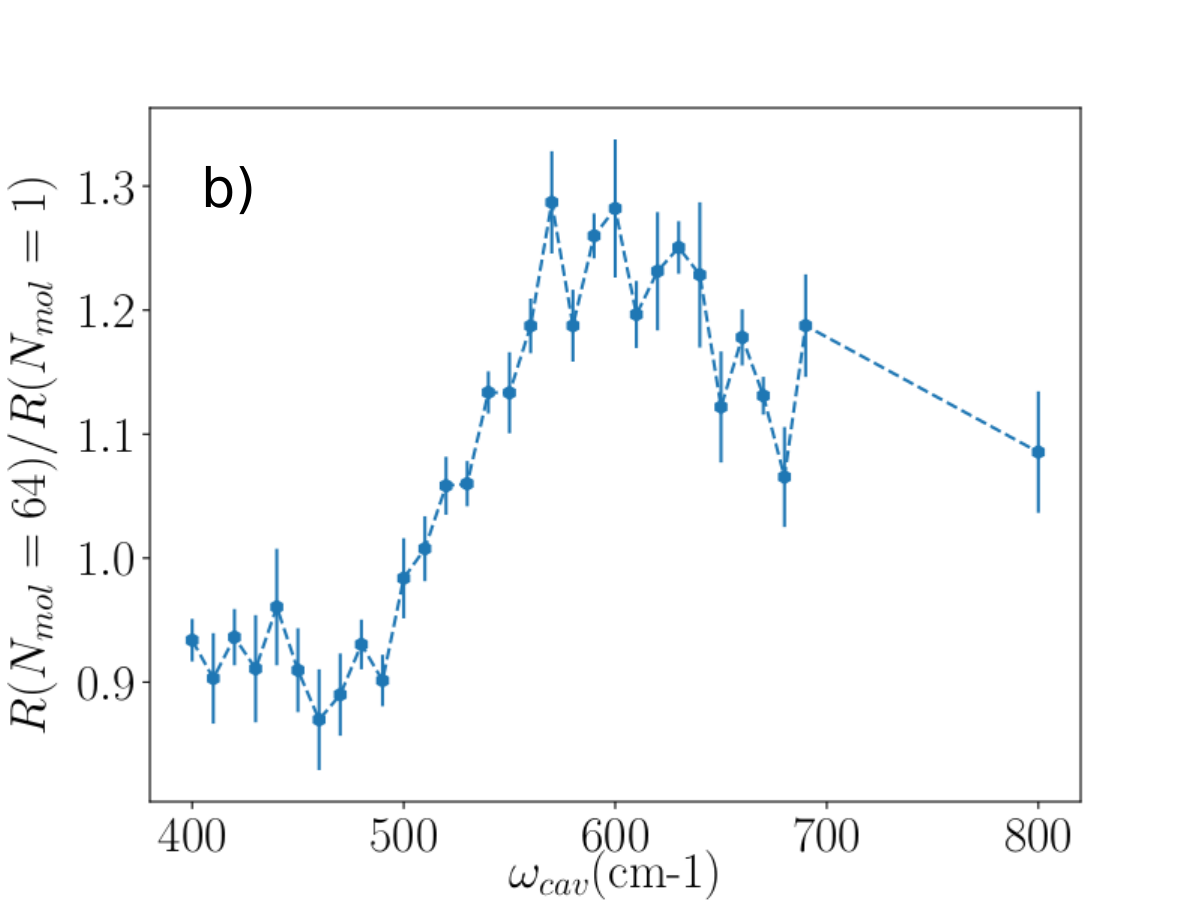} \caption{ \label{fig:R_tot} a) $R$
    for various cavity frequency from $400$ to $800$~cm-1. $g=0.4$ V/nm and
    $T=300$ K for all calculations. b) A ratio of $R(N_{mol}=64)$ and
    $R(N_{mol}=1)$ }
\end{figure}
Once we have gained an understanding of the role of molecular orientation the
question arises, through what specific mechanism do the NAMs participate in the
modification of the chemical rate.
Figure~\ref{fig:R_tot} shows $R$ as a function of $\omega_{cav}$ from $400$~cm-1
to $800$~cm-1 ($R$ in the $\omega_{cav}$ range $200$~cm-1 to $4000$~cm-1 is
provided as SI).
Interestingly, the rate modification peaks at a different frequency for the
cases $N=1$, $\approx$460~cm$^{-1}$, and $N=64$, $\approx$520~cm$^{-1}$, with the single molecule
case (necessarily the AM) being red-shifted by roughly 60~cm$^{-1}$.
Considering the ratio of $R(N=64)/R(N=1)$ in Fig.~\ref{fig:R_tot}b, one sees that the maximum collective
effect peaks at $\omega_{cav}\approx 600$~cm$^{-1}$.
These differences hint at the fact that the resonance condition between the
cavity and the AM, which is the decisive element by which the rate can be
modified, is modulated by the presence of the NAMs.

\begin{figure}[!h]
\includegraphics[width=8cm]{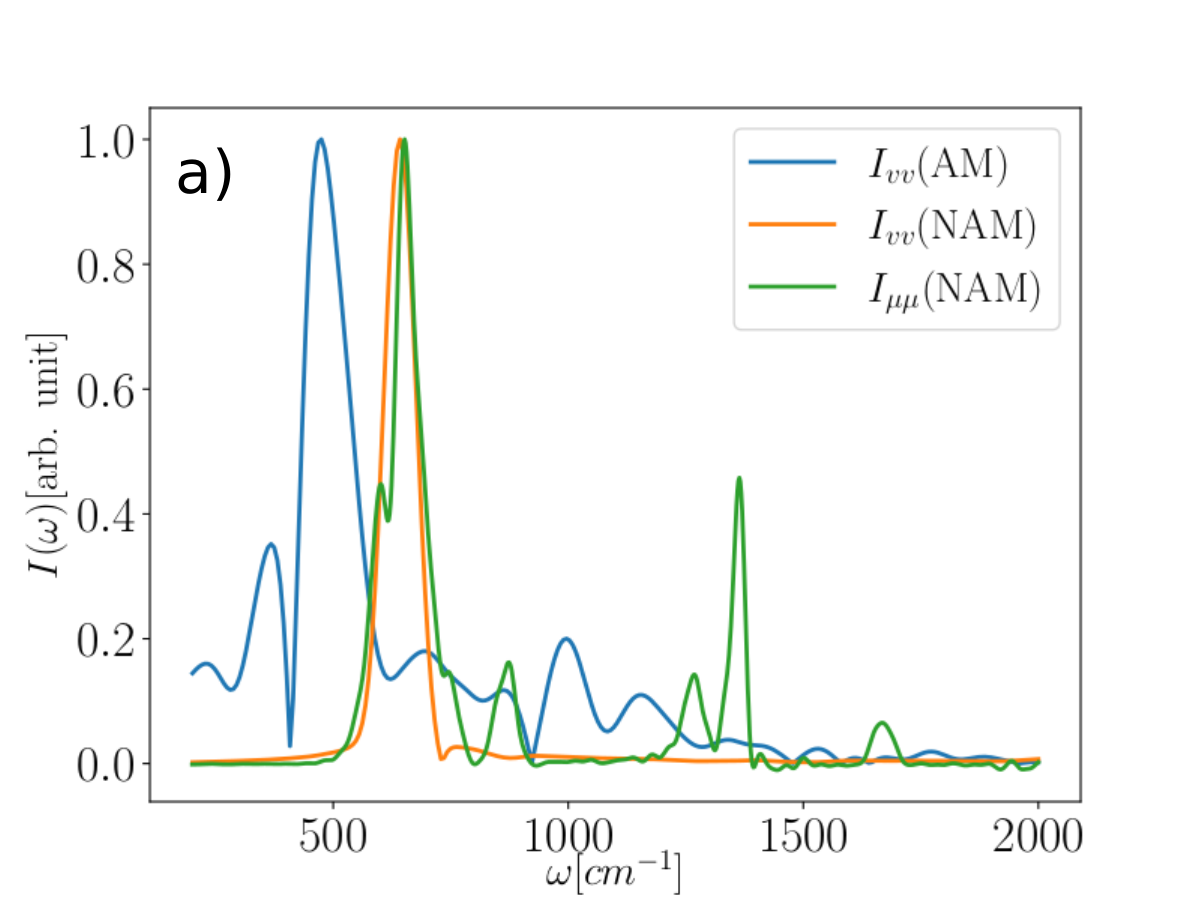}\\
\includegraphics[width=8cm]{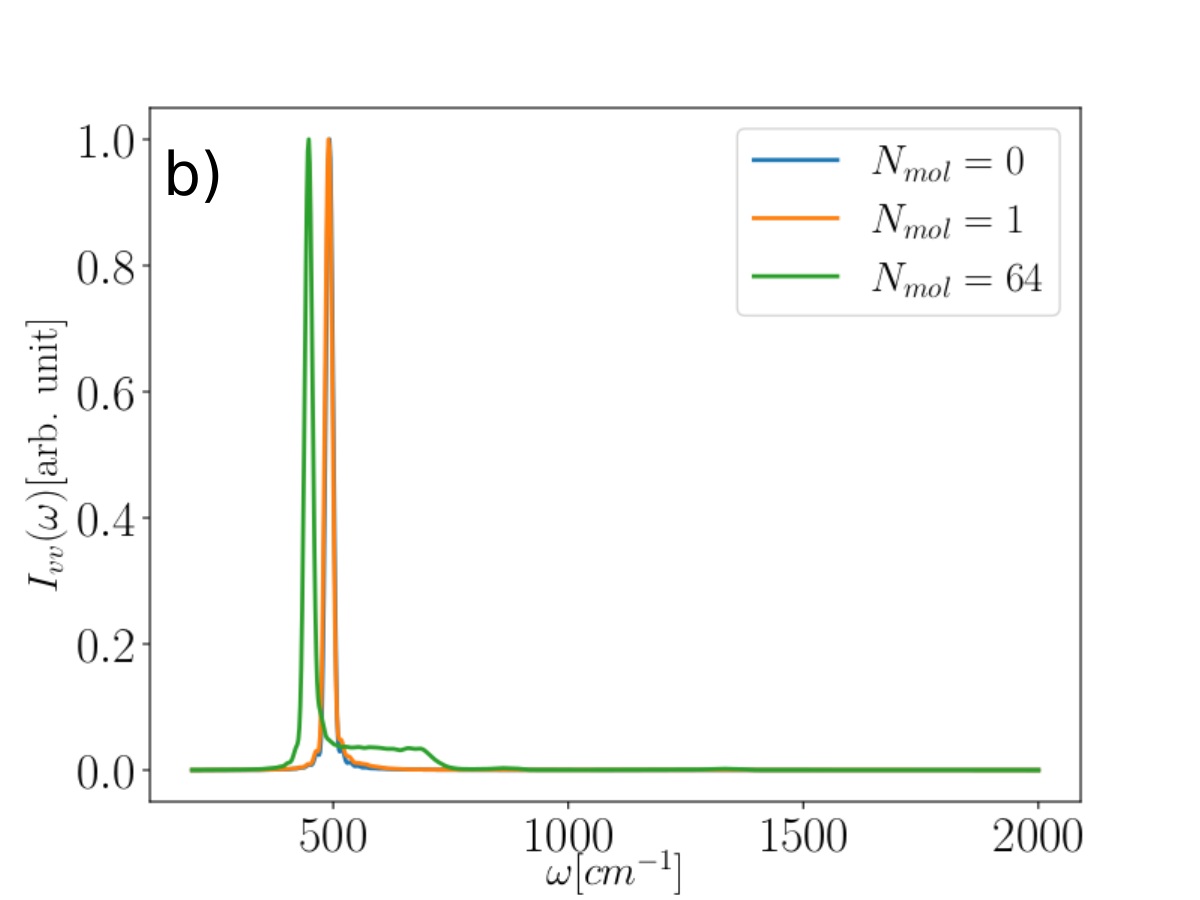}\\
\includegraphics[width=8cm]{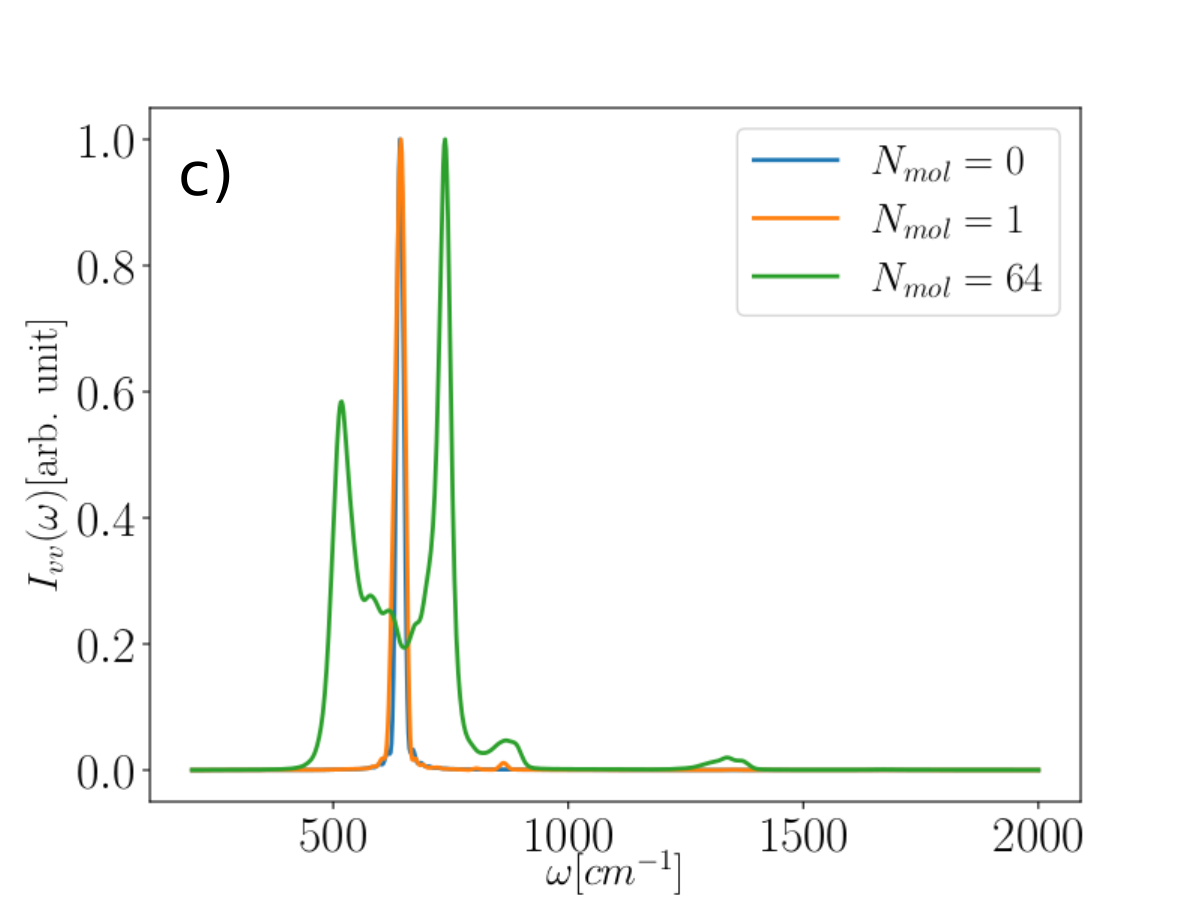}
\caption{
    \label{fig:spec}
     a) IR spectrum of non-activated HONO(NAM) and activated HONO(AM). The spectra are shown in absolute value.
     b) Velocity-velocity correlation spectrum of the cavity mode for $\omega_{cav}=490$~cm$^{-1}$.
     c) Same as b) for $\omega_{cav}=640$~cm$^{-1}$.
     $g=0.4$~V/nm and $N_{mol}=0, 1, 64$.
 }
\end{figure}

\change{change from here}
To shed light on this matter,
 Fig.~\ref{fig:spec}a compares the spectrum
 of the velocity-velocity correlation function of the reaction
 coordinate, $I_{vv}(\omega)$, for AM and NAM i.e. for HONO trajectories
 sampled from a thermal distribution at the TS dividing surface,
 and for HONO molecules in thermal equilibrium
 around the \emph{cis} configuration, respectively.
 Fig.~\ref{fig:spec}a shows that the vibrational frequency of the AM
 along the reaction coordinate is red-shifted compared to the NAM,
 which is
 caused by the anharmonicity of the potential well along this coordinate and
 the higher energy content of the AM.
Quantum mechanically, the anharmonic vibrational energy levels with the excitation
localized along the reaction coordinate become closer, in particular
\mbox{$\tilde{\nu}_{01}=632$},
\mbox{$\tilde{\nu}_{12}=581$},
\mbox{$\tilde{\nu}_{23}=555$},
\mbox{$\tilde{\nu}_{34}=515$}, and
\mbox{$\tilde{\nu}_{45}=467$}~cm$^{-1}$~\cite{ric_04_1306, ric_07_164315}, which classically results in a longer
oscillation period at higher energy.
The velocity-velocity spectrum of the NAM can be compared with the
spectrum of the dipole-dipole
correlation function~\cite{mcq_00_,nit_06},
$I_{\mu\mu}(\omega)$, confirming that the peak at about 600~cm$^{-1}$
in $I_{vv}(\omega)$ corresponds to the vibrational frequency of the NAM
along the reaction coordinate.
To validate the classical spectra we compare to an anharmonic quantum mechanical
spectrum calculated with the MCTDH approach using the Heidelberg package~\cite{Meyer1990,bec_00_1},
provided in the Supporting Material and showing qualitatively good agreement.

The blue-shifted vibrational frequency of the NAMs compared to the AM does not
explain by itself the fact that, as the number of molecules increases, the
largest rate modification occurs toward higher cavity frequencies.
First,
the AM should move off-resonance as $\omega_{cav}$ shifts to the blue.
Second, the NAMs are not undergoing the chemical reaction, so the fact
that their resonance condition improves with
a blue-shifted cavity should, in principle, not change the transmission
coefficient of the AM.
%
%
Figures~\ref{fig:spec}b and \ref{fig:spec}c show the spectrum of the
velocity-velocity correlation function, $I_{vv}(\omega)$, for the cavity mode
for $\omega_{cav}=490$ and $640$~cm$^{-1}$, respectively, and for 0, 1 and 64
molecules in the cavity.
At $\omega_{cav}=490$~cm$^{-1}$ and $N=1$, $I_{vv}(\omega)$ presents a single peak at
490~cm$^{-1}$ in all cases, without the formation of polaritonic peaks due to
the relatively small single-molecule coupling, and with just a small red-shift
for $N=64$ (green trace in Fig.~\ref{fig:spec}b).
On the other hand, and in stark contrast, $I_{vv}(\omega)$ presents two
polaritonic peaks for the $N=64$ case and $\omega_{cav}=640$~cm$^{-1}$.
Indeed, we want to emphasize that the single-molecule coupling
strength is set such that no polaritonic branches exist for one single
molecule, but such that they develop while increasing the number of molecules at fixed
coupling from $N=1$ to $N=64$.
%
Hence, at $\omega_{cav}=640$~cm$^{-1}$ the cavity is resonant with the NAMs and
the collective coupling is strong enough to split the cavity frequency in an
upper (UP) and lower polaritonic (LP) resonance. The LP is resonant with the
red-shifted frequency of the AM, which leads to an efficient coupling and to the
corresponding modification of the rate.

Thus, the following conclusions can be drawn:
(1) For an effective modification of the rate in the single-molecule, strong
coupling regime, the AM must be efficiently coupled and resonant with the
cavity. This can be through the reaction coordinate directly, or through other
modes anharmonically coupled to the reaction coordinate, as we have demonstrated
previously~\cite{jin_22_4441}.
(2) In the collective $N>>1$
strong-coupling regime,
the AM must be resonant with vibrational modes of the cavity dressed by the
NAMs, meaning for example resonances of the AM with polaritonic modes
formed by the $N-1$ spectators and the cavity.

\begin{figure}[!h]
\includegraphics[width=8cm]{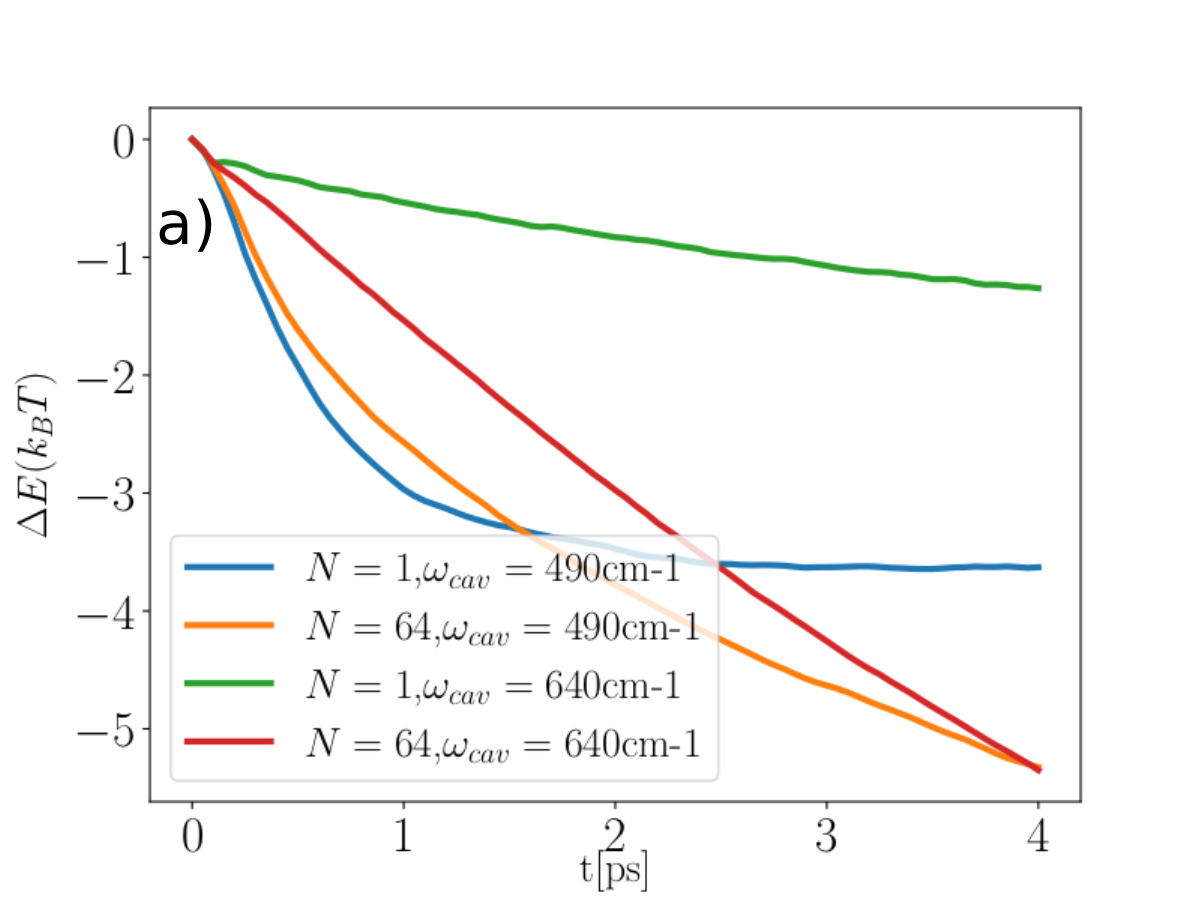}
\includegraphics[width=8cm]{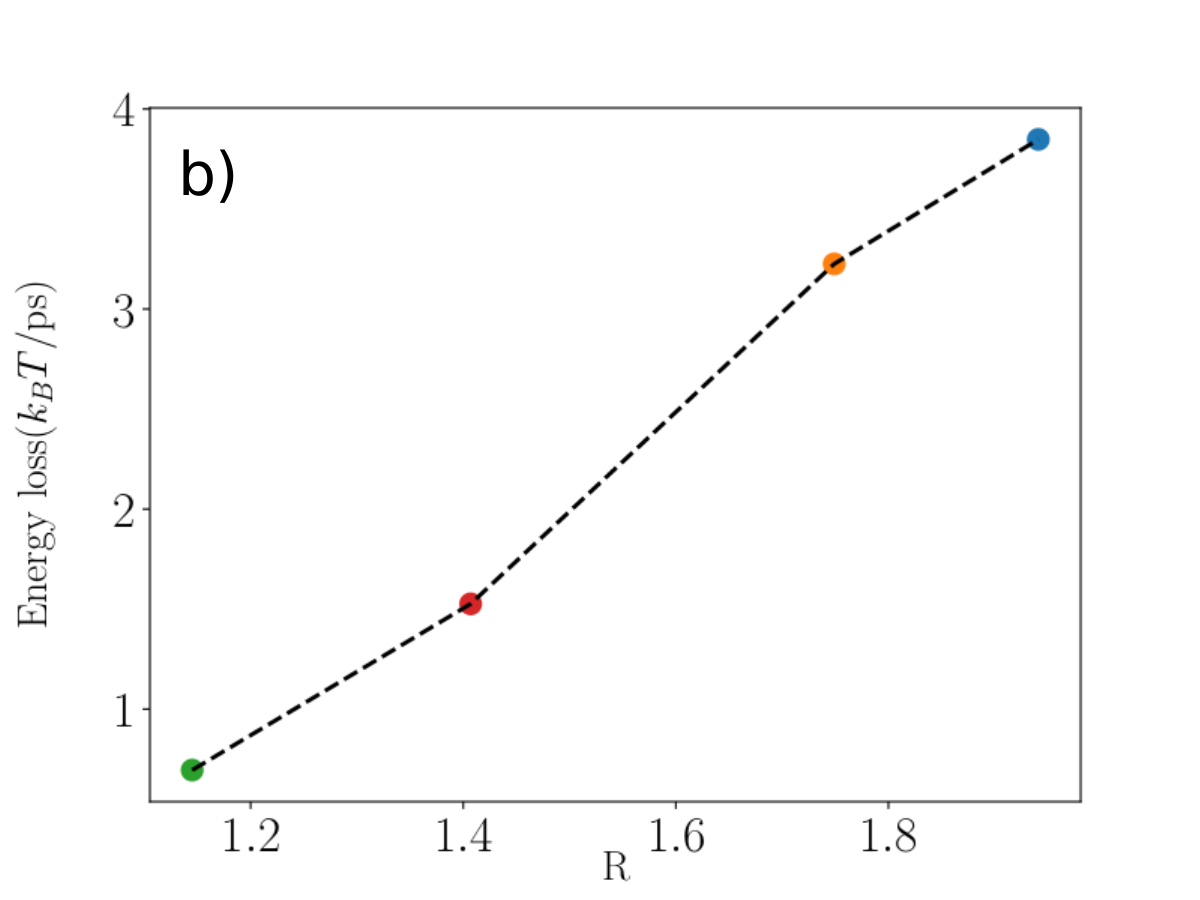}
\caption{
    \label{fig:E_dist}
     a) The energy difference,
     $\Delta E=E(t)-E(0)$, of AM as a function of time is plotted with $g=0.4$ V/nm and $T=300$ K for all
     calculations. b) The average energy loss, $E(t)-E(0)/t$, of AM within the first 0.5~ps is plotted as a function of R. The color of each point corresponds to the conditions shown in a).
    }
\end{figure}

One can still ask the question, what is the underlying physical mechanism
by which the cavity plus NAMs change the reaction rate of AM.
For this, we focus on the total energy variation of the AM after it
has passed the transition state, $\Delta E=E(t)-E(0)$, as a function of time
as plotted in Fig~\ref{fig:E_dist}.
The total energy of the cavity plus all molecules
is divided into three subsystems, the AM, the NAMs and the cavity, and
we consider the case for random molecular orientation.
The energy barrier is about 4000~cm$^{-1}$ or $20$~$k_{B}T$ at $T=300$~K.

The fastest energy loss from the AM occurs for $N=1$ and $\omega_{cav}=490$~cm-1, closely
followed by $N=64$ and $\omega_{cav}=490$~cm-1 (blue and orange traces). The corresponding
relative change in rate is $R=1.9$ and $R=1.8$, respectively (cf. Fig.~\ref{fig:R_tot}b).
The energy loss from the AM is slower at $\omega_{cav}=640$~cm-1 than at
$\omega_{cav}=490$~cm-1. At $\omega_{cav}=640$~cm-1, the $N=64$ case loses AM energy
faster than $N=1$, since in the latter situation the resonance
of the AM is largely detuned from the cavity, whereas for $N=64$ the AM is resonant with the
lower polaritonic state. The corresponding relative changes in the reaction rate are
$R=1.4$ and $R=1.2$ for the $N=64$ and $N=1$ cases, respectively.
Thus, the relative change in rate and the energy loss from the AM at short times are directly
correlated.
The better the AM is resonant with a mode of the cavity plus NAMs, the fastest the
energy exchange of the AM with the rest of the system at short times after crossing the TS,
and the largest the relative change in the chemical reaction rate.
Only the energy transfer at short times is relevant, since $\kappa^{(c)}(t)$ reaches its
plateau value when the AM becomes, in average, trapped in the reactant or product wells
due to energy loss.

     %



Summarizing, we have investigated the effect of a cavity mode coupled to an
ensemble of randomly oriented HONO molecules in the gas phase on the rate of the
\emph{cis}-\emph{trans} isomerization reaction.
Our simulations demonstrate that the activated molecule, the molecule undergoing
the chemical reaction at a given instant of time, and the ensemble of
non-activated molecules, play fundamentally different roles.
Specifically, the orientation with respect to the cavity that leads to the
largest effect can be different for the activated molecule and the non-activated
molecules. This is a consequence of the possible change of direction of the
permanent dipole in real space for different configurations of the
reacting molecule along the reaction coordinate in realistic systems.
Also, the relevant resonances of the activated molecule that couple with the
cavity may be shifted with respect to those of the non-activated molecules due
to the much larger energy content of the former.

The largest modulation in the transmission coefficient occurs when the reaction
coordinate (or a mode strongly coupled to the reaction coordinate~\cite{jin_22_4441})
in the activated molecule becomes resonant with the cavity or a polaritonic resonance
of the cavity with the non-activated molecules.
This does \emph{not} mean that the polaritonic resonance is already populated at
room temperature before the chemical reaction takes place. It means that the
polaritonic excited states can be populated through coupling to
the activated molecule.
Finally, the main mechanistic cause for the relative change in the transmission
coefficient, $R=\kappa^{(c)}/\kappa^{(0)}$, is the average rate of
energy loss from the
activated molecule at short times, i.e. until $\kappa^{(c)}$ reaches its plateau
value, which occurs due to the resonant coupling to the rest of the polaritonic
system.

Our simulations cover the single-molecule to small-ensemble strong coupling
regime in the gas phase, and therefore one should be cautious about
extrapolating our results to the macroscopic limit of large $N$ in Fabry-Perot
cavity experiments.
Strong coupling in the gas phase with methane molecules has recently been
observed by the Weichman's group~\cite{wri_02_185}, and this could represent an
interesting avenue for gas-phase experiments of cavity-modified molecular
reactivity under well-controlled conditions, where solid mechanistic insights
can be gained by the interplay of experiment and theory.
%
%
Our ﬁndings shed important new light onto the question of collective effects in
chemical reactivity under vibrational strong coupling. However, it still remains
for future work to better understand how these cavity eﬀects can survive in
actual liquid phases and in the collective regime for truly macroscopic numbers
of molecules.
We speculate that these answers might lie beyond the paradigm of
non-interacting molecules with idealized cavity modes, and unveiling them may
require studies of the transmission coeﬃcient with full consideration of the
molecular as well as electromagnetic environments.

\begin{acknowledgement}
    J.S. is grateful to the International Max Planck Research School for
    Quantum Dynamics in Physics, Chemistry and Biology (IMPRS-QD) for financial support.
    We acknowledge financial support by the Deutsche Forschungsgemeinschaft (DFG)
    through project 429589046.
\end{acknowledgement}

\begin{suppinfo}
    Additional theoretical and numerical details including extra figures.
\end{suppinfo}

\bibliography{zotero_fullbt_ov}

\end{document}